\documentclass[12pt]{iopart}
\usepackage{graphicx}
\begin{document}
\title[ Resonant growth of  stellar oscillations ]
{ Resonant growth of  stellar oscillations by 
incident gravitational waves }
\author{Yasufumi Kojima\footnote[1]{Email:
 \mailto{kojima@theo.phys.sci.hiroshima-u.ac.jp} }
and Hajime Tanimoto
}
\address{Department of Physics,
Hiroshima University,
Higashi-Hiroshima 739-8526, Japan}
\begin{abstract}
Stellar oscillation under the combined influences of
incident gravitational wave and radiation loss is studied
in a simple toy model.
The star is approximated as a uniform density ellipsoid
in the Newtonian gravity including 
radiation damping through quadrupole formula. 
The time evolution of the oscillation is significantly controlled
by the incident wave amplitude $h$, 
frequency  $\nu$ and damping time $\tau$.
If a combination  $ h \nu \tau $ exceeds a threshold value, 
which depends on the resonance mode, 
the resonant growth is realized.
\end{abstract}
%
\maketitle
\section{ Introduction }

  Stellar oscillation coupled to gravitational waves is
one of important issues for the gravitational wave 
observatories such as LIGO, VERGO, GEO600 and TAMA. 
The current status of these detectors is given 
elsewhere(e.g, \cite{gwd8}). 
Detection of gravitational radiation from compact stars 
would provide valuable information about their interior.
The frequency and damping time are uniquely
affected by the stellar structure. 
Gravitational wave asteroseismology in the future is also 
discussed(e.g, \cite{ak98, bfg04} ).
Various oscillation modes in 
neutron stars have been studied theoretically so far,
but little is known of the excitation and damping processes. 
Most likely event that may excite the oscillations is the birth 
of the neutron stars after supernova. 
Some oscillation modes may also be induced by instabilities.
An accurate description of these processes requires 
numerical solution of coupled system of hydrodynamics and
the Einstein equations. 
Recently, the numerical simulations have been developing 
in the several approaches,
e.g, 2D Newtonian simulation\cite{oblw04}, 
3D smooth particle code simulation\cite{fhh04},
Newtonian MHD simulation\cite{ketal04}, 
GR  simulation\cite{ss04}
and the referees therein.

  Approximate approach is also helpful to get some insight 
into the physical processes.
For example, excitation of neutron star oscillations 
in a close binary is studied  by using following 
approximation\cite{ko87, rlp01, tsm01, pbgmf02}.
One of the two stars is an extended body, and the other 
is approximated by a pointlike particle.
The stellar perturbations in general relativity
are solved for the extended body, while the pointlike mass 
moves on geodesic of the spacetime around the star.
The oscillation mode can be excited by the particle, whose orbit
is close enough to satisfy the resonance condition
between normal mode frequency of the star and
orbital Kepler frequency.
Detailed discussion  concerning excitation of neutron star
oscillation in this approach may be found in Ref.\cite{pbgmf02}
and references therein.
This is one of examples showing that oscillations are
excited by the external almost periodic disturbance.
The resonant growing mode is in general significantly
affected by viscous damping, which may come from 
physical origins or numerical truncation errors.
It is therefore better to examine  resonant property 
in a simplified model.
In this paper, we consider an idealized situation to
study the resonant behavior.
Incident gravitational waves are regarded as
external periodic disturbance. 
The purpose of our study is the following.
What is the resonance condition?
How does the amplitude grow?
The external disturbance is characterized by the
frequency, duration and amplitude of the waves.
These quantities are closely related with the resonance condition 
and the amplification of growing mode.
The resonance condition is a relation between 
intrinsic and external frequencies. 
The amplification factor depends on 
growth rate of the mode and duration of the disturbances. 
%

  This paper is organized as follows.
In section 2, we briefly summarize the resonance appeared
in a harmonics oscillator, which is helpful to examine
the stellar model.
In section 3, we consider the stellar oscillation 
with ellipsoidal model under Newtonian gravity. 
The gravitational radiation loss is included by the
quadrupole formula. 
We show the numerical results in section 4. 
Section 5 is devoted to discussion.

\section{ Resonance in a harmonic oscillator}

  We here briefly consider the resonance in a harmonic oscillator,
which is a useful model to understand the more realistic case below.
There are two kinds of resonance (e.g, \cite{llbook}).
One type occurs when the frequency $(=\Omega/2\pi)$
of external perturbation matches with the intrinsic 
frequency  $(=\omega/2\pi)$, i.e, $ \Omega = \omega$. 
The relevant equation is given by 
\begin{equation}
{\ddot X}  + \omega ^2 X =  f \cos(\Omega t) .
\end{equation}
In this case, the amplitude $X$ increases linearly with the time,
$ X \propto (f/2 \Omega) t \sin(\Omega t) . $
Another type of resonance is parametric resonance, which
occurs for time-dependent frequency.
One example known as the Mathieu's equation is 
\begin{equation}
{\ddot X}  + \omega^2 (1 + b \cos(\Omega t) ) X =0 .
\end{equation}
The exponential growth for instability appears when 
$ \omega / \Omega = n/2, ( n=1,2, 3, \cdots ).$
There is allowed range of frequency for each
resonance condition, but the width of resonance range 
decreases rapidly with increasing $n$.
The matching for large $n$ is hardly satisfied.
The strongest instability occurs for the fundamental one $n=1$.
Therefore the parametric resonance with $n=1$
is very important.
The growth rate $s$ in $ X \propto \exp( s t)$
for the fundamental mode
can be calculated for small $b $  as
\begin{equation}
 s = - \frac{1}{2}     
\left[ - \left( \Omega -2 \omega  \right)^2
       +  \frac{1}{4} b^2 \omega ^2    \right] ^{1/2}
\sim  \frac{1}{4} b  \omega   
\end{equation}
for the unstable region
\begin{equation}
 - \frac{1}{2} b \omega      < \Omega -2 \omega
< \frac{1}{2} b \omega   .   
\end{equation}
In the presence of frictional damping,
the amplitude decreases with time as  $ \exp(-\gamma t)$. 
The damping counteracts the parametric resonance,
so that the condition is more constrained.
The growing solution of the resonance 
is possible only when 
$b$ exceeds a threshold  $ 4 \gamma /\omega $.

\section{ Stellar pulsation }

 Now we consider the pulsation of the star based on Newtonian gravity.
The equation of motion for the pulsation 
driven by external acceleration  ${\vec g} $ is
\begin{equation}
\frac{d {\vec v}}{dt}  = - \frac{1}{\rho} \vec{\nabla} p
 -\vec{\nabla} \phi +{\vec g} .
\end{equation}
As an external perturbation, we consider plane parallel wave 
with amplitude $h$ and angular frequency $\Omega$,
propagating along $z$-axis.
The tidal acceleration by the incident gravitational wave
is expressed as\cite{mtw}
\begin{equation}
{\vec g} =  \left[ -\frac{1}{2} h \Omega ^2 x \cos\Omega(t-z),~ 
 \frac{1}{2} h \Omega ^2 y \cos\Omega(t-z),~ 
 0 \right]. 
\label{incgw}
\end{equation}
We here consider + mode only.
Another polarization mode $\times$ is given simply 
by $ \pi/4$ rotation about $z$-axis, and may be ignored
without loss of generality. 
Deceleration due to quadrupole radiation loss
is given by\cite{mtw}
\begin{equation}
{\vec g} =  - \vec{\nabla}
 \frac{G}{5c^5} \mathcal{I}_{ij} ^{(5)} x^i x^j ,
\label{gwloss}
\end{equation}
where $ \mathcal{I}_{ij} ^{(5)}$ is five time
derivatives of reduced quadrupole moment.
We incorporate both effects by eqs.(\ref{incgw}) and (\ref{gwloss}).

  We are interested in global motion of a star,
and therefore approximate it by the  ellipsoidal model.
In the approximation,
the dynamical motion is limited to uniform expansion 
and compression along three axes.
Thus, the fluid motion is described by three
functions of time $a_i(t) (i=1,2,3)$.
The ellipsoidal approximation is often used as
a simplified model to extract dynamical features in
more realistic cases.
(See e.g, \cite{ros, prte, mil, calu, larash}.)  
Assuming incompressible fluid with uniform density $\rho$,
the dynamics can be determined by
\begin{eqnarray}
\label{dyna1.eqn}
{\ddot a_1 }  &=& -2\pi G \rho a_1 A_1  + \frac{K }{a_1} 
-\frac{1}{2 }h  \Omega ^2 a_1\cos\Omega t
  - \frac{2G}{5c^5} \mathcal{I}_{xx} ^{(5)} a_1 ,
\\
{\ddot a_2 }  &=& -2\pi G \rho a_2 A_2 + \frac{K }{a_2} 
+ \frac{1}{2}h  \Omega ^2 a_2 \cos\Omega t
  - \frac{2G}{5c^5} \mathcal{I}_{yy} ^{(5)} a_2 ,
\\
{\ddot a_3 }  &=& -2\pi G \rho a_3 A_3  + \frac{K }{a_3} 
- \frac{2G}{5c^5} \mathcal{I}_{zz} ^{(5)} a_3,
\label{dyna3.eqn}
\end{eqnarray}
where the function $A_i$ depends on the
shape of the ellipsoid, and is described in \cite{ch69}.
The term $K$ associated with pressure
is algebraically determined from the condition
$ d( \ln (a_1 a_2 a_3 ) )/dt =0.$
The system of eqs.(\ref{dyna1.eqn})-(\ref{dyna3.eqn})
depends on three dimensionless parameters, 
$ Q = \omega \tau /2\pi = \nu  \tau, ~\omega / \Omega$ and $ h,$
where $  \nu (=\omega/2 \pi) $ is intrinsic frequency of
the stellar oscillation, and $ \tau $ is damping time due to
gravitational radiation.
The amplitude becomes $e^{-1}$ after $Q$ times oscillations. 
These values are physically related with mass $M$
and radius $ R$  of the star as
\begin{equation}
  \omega = \left( 
\frac{ 4GM}{5 R^3} \right) ^{1/2},
~~~~
  \tau ^{-1} =  
\frac{ 2GM R^2 \omega ^4}{25 c^5} .
\end{equation}
%


In order to clarify the dynamics of this system, we
assume small deviation from spherically symmetric 
equilibrium ($a_i =1$), and linearize 
eqs.(\ref{dyna1.eqn})-(\ref{dyna3.eqn}).
We neglect the radiation loss for simplicity and  
limit the oscillation  to the toroidal motion in $x$-$y$ plane.
The dynamics can be expressed by a single function
$X$  as
\begin{equation}
{\ddot X }  +\omega ^2 
\left( 1 + \frac{h}{2}
 \left(\frac{\Omega}{\omega} \right) ^2 \cos \Omega t 
\right)X  = - \frac{1}{2} h \Omega ^2 \cos \Omega t,
\label{linear.eqn}
\end{equation}
where $ X$ represents the deviations $ \delta a _i (= a_i -1)$ 
from the equilibrium state:  
\begin{equation}
  \delta a_1 = -  \delta a_2  = X,
~~~
\delta a_3 =0 .
\end{equation}
Equation (\ref{linear.eqn}) is the Mathieu's one
with periodic source term.
As considered in section 2,
it is clear that two kinds of resonance are possible in 
eq.(\ref{linear.eqn}), and hence in the system of eqs.  
(\ref{dyna1.eqn})-(\ref{dyna3.eqn}).
In the astrophysical situation, 
periodic external perturbation may be expected, 
e.g, in close binary system, where
the standard resonance due to the periodic driving force
is relevant.
On the other hand, we do not definitely know the 
situation in which the parametric resonance occurs, but
the relevant elements may be involved in
the complicated system like  the hydrodynamical core collapse.

\section{Numerical calculation}

%
\subsection{ Off-resonant case  }
Before examining the resonant properties of
the system (\ref{dyna1.eqn})-(\ref{dyna3.eqn}), we first
of all show the behavior of non-resonant case.
This is necessary for checking and comparing models.
  As for initial conditions of numerical calculations, 
we adopt static state, i.e,
($a_1 = a_2 = a_3 =1,  \dot  a_1 = \dot a_2 = \dot a_3 =0$).
The oscillation amplitudes $ \delta a_1 $, $ \delta a_3 $
as a function of time are shown in Fig.1. 
The amplitude $ \delta a_2 $ is always well 
approximated as $ \delta a_2  \approx - \delta a_1 $,
and is omitted in the figure.
The parameters adopted for the calculation are
$ Q = 90$, $ \omega / \Omega = 0.6 $ and $ h  =0.1$.
The damping time $\tau$ of the model corresponds to
$ \tau =2 \pi Q/ \omega =  150 \times (2 \pi /\Omega )$.
In the early stage of the oscillation,
the temporal profiles are a mixture of two 
modes with frequencies 
$ \omega (= 0.6 \Omega)$ and $ \Omega $. 
There is a modulation of two modes, which 
causes  larger amplitudes of the oscillations
due to interference of two waves. 
As the time goes on, 
the amplitudes of both modes are damped by the radiation loss. 
Since the external perturbation 
with  $ \Omega $ is always supplied,
the mode becomes dominant around  $\tau  $
$(= 150 \times (2 \pi /\Omega )  )$, 
and the oscillation is eventually enforced to match it
at the late phase, typically at several times $ \tau $,
as shown in right panel of Fig.1.
From the numerical results, we found that
the motion along $a_3$ is also induced by 
the incompressible condition 
$ a_3 =1/(a_1 a_2 )$.
This differs from linear analysis, in which
$ a_3 $ is always constant.
The oscillation of $ \delta a_3 $ is of order $ h^2 $ 
in the amplitude, 
and is induced by a mixture of stellar intrinsic 
oscillation, $ \omega $ and external perturbation, $\Omega $. 
The frequencies of $\delta a_3$ are therefore described by
$ 2 \Omega$, $ 2 \omega$  and $ \Omega \pm \omega$.
%
\begin{figure}[ht]
\begin{center}
\includegraphics[scale=1.0]{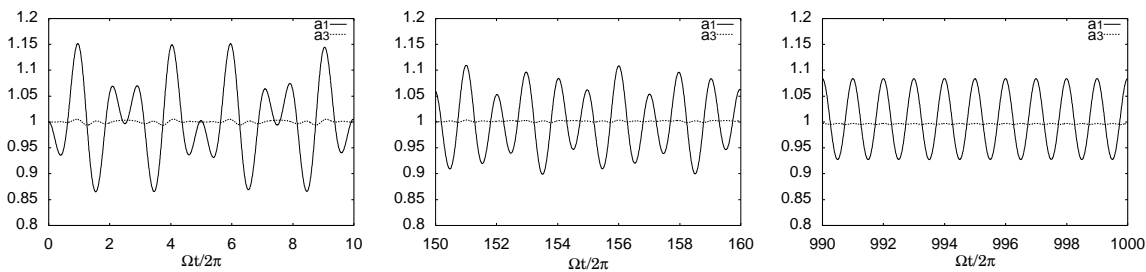}
\end{center}
\caption{
Oscillation amplitudes $\delta a_1$ and  $\delta a_3$ 
as a function of time $ \Omega t /(2\pi)$.
Parameters of the model are $Q=90$,
$ \omega / \Omega =0.6$ and  $ h=0.1  $.
From the left to right panels, 
the amplitudes of the oscillations are
shown for the early, middle and late stages.
}
\end{figure}

%
  We have also calculated the oscillations for
different parameters, and found that
the general behavior is almost the same
unless the resonance condition is satisfied.
The temporal behavior may be easily understood by analyzing
the numerical data with  Fourier transformation. 
It is found that the oscillation is approximated by 
the following curve consisted of two frequencies
$ \omega $  and $ \Omega $:   
\begin{equation}
\delta a_1 \approx 
- \delta a_2  \approx 
  \frac{ \Omega ^2 h}{2 |\Omega^2 - \omega ^2 |} 
\left( \cos \Omega t -
\exp( -t/ \tau ) \cos \omega t \right),
~~~~
 \delta a_3  \propto  h^2 .
\end{equation}
This formula can be derived from the linearized 
system eq.(\ref{linear.eqn}) 
and the accuracy is checked by fitting 
the numerical results for a wide range of parameters.   

In the limit of $ t \to \infty $,
the oscillation with  $ \omega $ in $a_1$ or $a_2$ 
is completely damped, and  
the mode matched with the incident wave survives
with the amplitude $ \Omega ^2 h /(2 |\Omega^2 - \omega ^2 | )$.
The amplitude of the steady state realized 
at $ t \to \infty $ becomes larger when $\Omega \to  \omega $.
The amplitude apparently diverges at the resonance, but
does not actually, because even small dissipation is
important and suppresses the divergence in that case. 
See section 4.3.

\subsection{ Parametric resonance of fundamental mode }
  In this section,  we consider the parametric resonance
of the fundamental mode by setting  $\omega / \Omega =1/2$.
In Fig.2, we show time evolution of the model with
$ h = 0.1, Q=45$.
The damping time $\tau$ corresponds to
$ \tau = 90 \times(2 \pi /\Omega )$.
In the early phase of the time evolution,
the amplitude of intrinsic stellar oscillation mode 
$ \omega (= \Omega /2 )$
exponentially grows up. After a hundreds of the oscillations,
damping effect becomes important around $ t \sim \tau$.
Eventually the mode $\omega $ is damped, and oscillation is 
enforced to external mode with  $ \Omega $ at the late
phase $ t \gg \tau$, in which 
the amplitude of  $ \delta a_1 $ becomes
$ \delta a_1 \approx 2h /3 . $
This behavior in the final steady state 
is the same as that in off-resonant case.
Compare the right panel in  Fig.1 
($ t \approx 1000 \times(2 \pi /\Omega )$)
with that in Fig.2 
($ t \approx 700 \times(2 \pi /\Omega )$).
Both figures show the behavior corresponding
to  several times $ \tau $.
Thus, the parametric excitation decays away in this case.
%

\begin{figure}[ht]
\begin{center}
\includegraphics[scale=1.0]{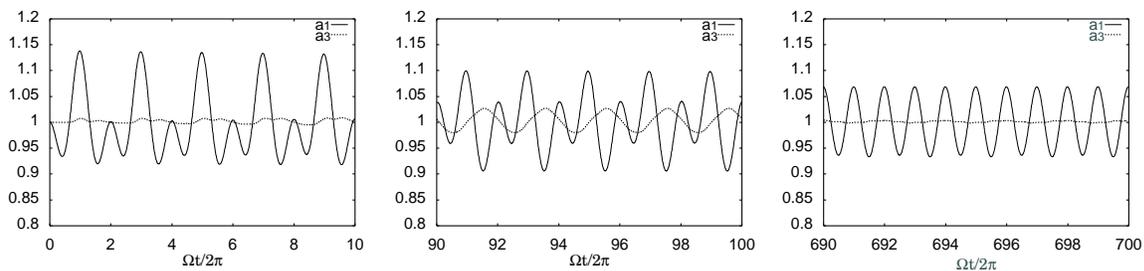}
\end{center}
\caption{
Oscillation amplitudes $\delta a_1$ and  $\delta a_3$ 
as a function of time $ \Omega t /(2\pi)$.
Parameters of the model are $Q=45$,
$ \omega / \Omega =0.5$ and  $ h=0.1  $.
}
\end{figure}

  We consider a case in which the 
radiation loss is less effective. 
The results for the parameters $ h = 0.1, Q=75$ 
are shown in Fig.3. The damping time $\tau$ corresponds to
$ \tau = 150 \times (2 \pi /\Omega )$.
In the early stage, a mixture of two frequencies
$ \omega (= \Omega /2)$ and $ \Omega $ can be seen.
There are no drastic differences between the left panels 
in Fig.2 and Fig.3 as for the early stages $ t < \tau $.
By comparing the right panels in Fig.2 and Fig.3, 
it is found that the parametric excitation occurs
in Fig.3, since the oscillation amplitudes 
of both $ \delta a_1$  and  $ \delta a_3$ are indeed enhanced.
The dominant oscillation frequency at the later stage
is determined by 
not external one $  \Omega $, but intrinsic one $  \omega $.
The frequency of $\delta a_3$ is also $ \omega $, which
comes from quadratic coupling between $\Omega$ and $ \omega$,
as $ \Omega -\omega ~(= \omega)$.
The other higher frequencies such as
$ 2\omega$, $ \Omega +\omega ~(= 3\omega)$ and 
$ 2\Omega ~(= 4\omega)$ are less excited. 
The amplitude at the later stage is saturated around 
the value $ \delta a_1 \approx \delta a_3 \approx  4 h $,
which is several times larger 
than that of off-resonant oscillation.
This saturation property is quite different from the 
linear theory, in which the amplitude
of the unstable mode exponentially increases so
far as resonant condition is satisfied.
In the nonlinear ellipsoidal model,
the motion along the longitudinal direction $ a_3$
plays a key role on the saturation as explained below.
The amplitudes of $ \delta a_1$ and $ \delta a_2$ grow
in proportion to $h$, while that of $ \delta a_3$, 
which is formally $ \sim  h^2$, also grows 
due to nonlinear coupling.
The amplitude of $\delta a_3 $ is given by
$\delta a_3 \sim (\delta a_1 )^2$.
When the oscillation amplitude becomes large enough,
say $ \delta a_1 \sim 0.5$,
$\delta a_3 $ is also large enough.
The motion of $\delta a_3$ is no longer neglected and
significantly affects the dynamics of $ \delta a_1 $ 
and $\delta a_2$.
In presence of the dynamics of $\delta a_3$
the growing amplitudes are saturated around 
the same order
$ \delta a_1 \sim \delta a_2 \sim \delta a_3 \sim 0.5$.

\begin{figure}[ht]
\begin{center}
\includegraphics[scale=1.0]{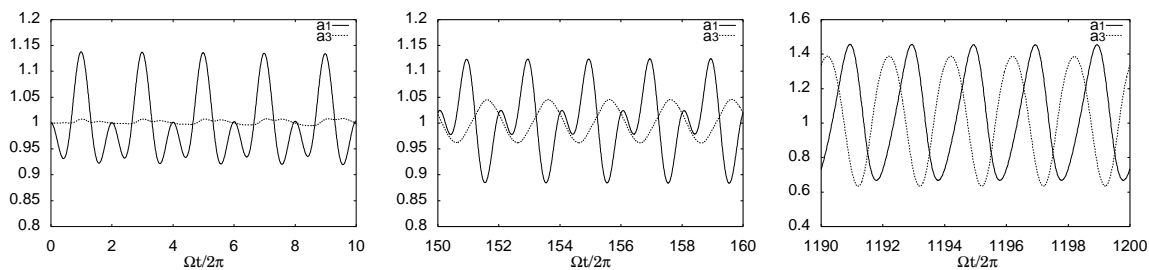}
\end{center}
\caption{
Oscillation amplitudes $\delta a_1$ and  $\delta a_3$ 
as a function of time $ \Omega t /(2\pi)$.
Parameters of the model are $Q=75$,
$ \omega / \Omega =0.5$ and  $ h=0.1  $.
}
\end{figure}

 So far we have shown two different results of the 
resonance in Fig.2 and Fig.3.
The parametric oscillation is excited 
for weak damping case, i.e, large $Q$, while it decays out
for strong damping case, i.e, small $Q$.
The excitation condition should also depend on the amplitude $h$
of the incident gravitational wave.
In order to  study the resonance condition 
for $\omega / \Omega =1/2$,
we have performed the numerical calculations
of eqs.(\ref{dyna1.eqn})-(\ref{dyna3.eqn})
for  various parameters $ (h, Q )$, 
typically in the range $ 20 < Q < 200$.
The results are summarized in Fig.4.

  We found that a combination of the parameters, $  h Q $
is a very important factor to determine 
the evolution\footnote{ 
Note that  $  h Q $ corresponds to $ b \omega /\gamma $ in the
harmonics oscillator shown in section 2.
The importance of this factor is reasonable.
}.
As shown in Fig.4, there is a critical curve discriminating 
between growth and damped cases. The 
curve is empirically given by  $ h Q \approx 7$.
Below the critical curve, the damping effect is
so strong that the parametric excitation is suppressed
like in Fig.2.
On the other hand, above the critical curve, the damping effect is 
less effective. In this case, 
parametric excitation of the intrinsic oscillation with
$ \omega $ occurs and the amplitudes can grow up.
Eventually,  the nonlinear effect becomes important and
the amplitudes of the unstable oscillations are
saturated around finite values.

\begin{figure}[ht]
\begin{center}
\includegraphics[scale=0.8]{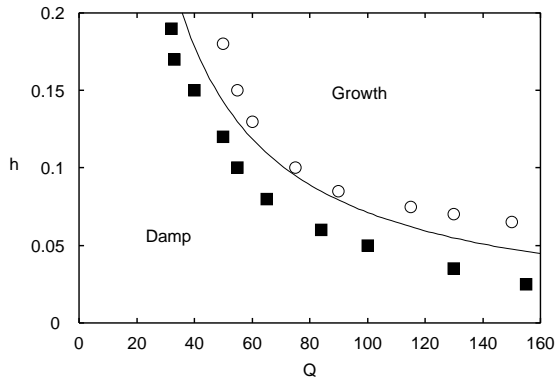}
\end{center}
\caption{
Excitation of the resonant oscillation significantly
depends on the 
parameters, $ Q $ and $h$.
Open circles and filled squares respectively denote
growth and damped cases in the final steady states. 
The critical curve discriminating two states
is approximately given by $ h Q \approx 7$. 
}
\end{figure}

\subsection{ Resonance  at $\omega =\Omega $ }

In this section, we consider the resonant behavior at
$\omega =\Omega $.
Two typical cases  are shown in Fig.5 and Fig.6.
These figures show their long term evolutions up to
hundreds of cycles of the oscillations,
so that each sinusoidal curve is not clear, but
the gradual change of the envelope is seen.
The  model parameters  of Fig.5 are 
$ Q=100$,  $h=5 \times 10^{-3}$, 
while those of Fig.6 are $ Q=100$,  $h=5 \times 10^{-2}$.
The behaviors are the same only for the 
initial stages, in which the amplitude of $ \delta a_1$
(or $\delta a_2$) increases linearly with time:
$ \delta a_1 \approx h \Omega t \sin(\Omega t)/4 $,
so far as $ \delta a_1 \ll 1 $.
The later behaviors significantly depend on the 
adopted parameters, and are exhibited in quite different way.
We have also calculated in a wide range of parameter space 
$(Q, h)$ and found that the most important 
factor is $ hQ$ like in the previous section.
The reason is explained below.

The oscillation of $\delta a_3 $ is inevitably
induced by the nonlinear coupling: 
$ \delta a_3  \propto \delta a_1 \delta a_2 \sim
( h \Omega t  )^2 \sin^2(\Omega t) /16$.
The amplitude of $\delta a_3 $ grows quadratically with time, and
is therefore small only in the initial evolution.
There is an epoch in which 
both amplitudes of $\delta a_1  $ and  $\delta a_3 $
become the same order. This non-linear timescale 
$ t_n $ is estimated as $ t_n = 4/(h \Omega) $
by setting $\delta a_1 = \delta a_3   $,
if the damping may be neglected.
It is important to compare 
this timescale with that of the radiation loss 
$\tau = Q \times 2 \pi /\Omega $.
If $ \tau < t_n  $, which corresponds to  
$ hQ  < 2/ \pi \approx 0.6$,
the radiation loss controls the evolution.
As shown in Fig.5 ($ hQ  = 0.5$), 
the grown amplitude of $\delta a_1 $ is saturated 
around the value 
$\delta  a_1 \sim h \Omega \tau  \sim h Q $.
The amplitude of $\delta a_3 $ grows up in much
longer nonlinear timescale, which is not
exactly but roughly given by $ t_n $.
We have shown in Fig.5 the time evolution for 
$ t < 250 \times (2 \pi/\Omega) $, but
confirmed that the subsequent evolution
for $ t > 250 \times (2 \pi/\Omega) $
becomes almost steady state with some small modulation
in the amplitudes.

On the other hand, 
if $ \tau > t_n  $, i.e,   $ hQ  >2/ \pi $,
the nonlinear coupling is important.
The behavior in this case is highly 
unstable as shown in Fig.6 ($ hQ  =5 $).
The amplitudes $ \delta a_i ~(i=1,2,3)$ become 
$ \mathcal{O} (1)$.
The nonlinear timescale in this model corresponds
to $ t_n \sim 13 \times  (2\pi / \Omega)$, where
the amplitude  $ \delta a_3$ becomes large.
The time evolution exhibits {\it chaotic behavior}\footnote{
In general, it may be necessary to measure e.g,
Lyapunov exponent in order to judge whether the
nonlinear behavior is chaotic or not. We did not 
calculate the exponent, but found that 
the numerical behavior is very sensitive like chaotic system.
},
and therefore 
the actuate numerical integration 
is rather difficult for much longer time, say,
up to $ 10^3 \times (2 \pi / \Omega)$ in Fig.6.

\begin{figure}[ht]
\begin{center}
\includegraphics[scale=0.8]{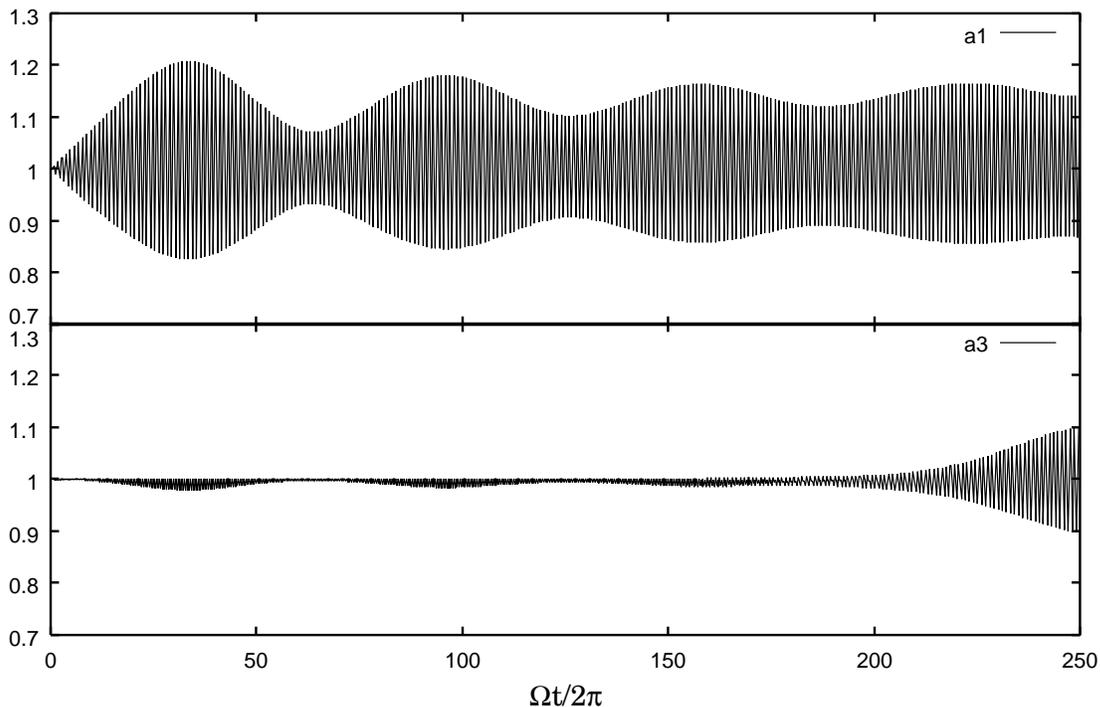}
\end{center}
\caption{
{
Oscillation amplitudes $\delta a_1$ and  $\delta a_3$ 
as a function of time $ \Omega t /(2\pi)$.
Parameters of the model are 
$ Q=100$, $ \omega =\Omega$ and  $h=5 \times 10^{-3}$ 
}  
}
\end{figure}

\begin{figure}[ht]
\begin{center}
\includegraphics[scale=0.8]{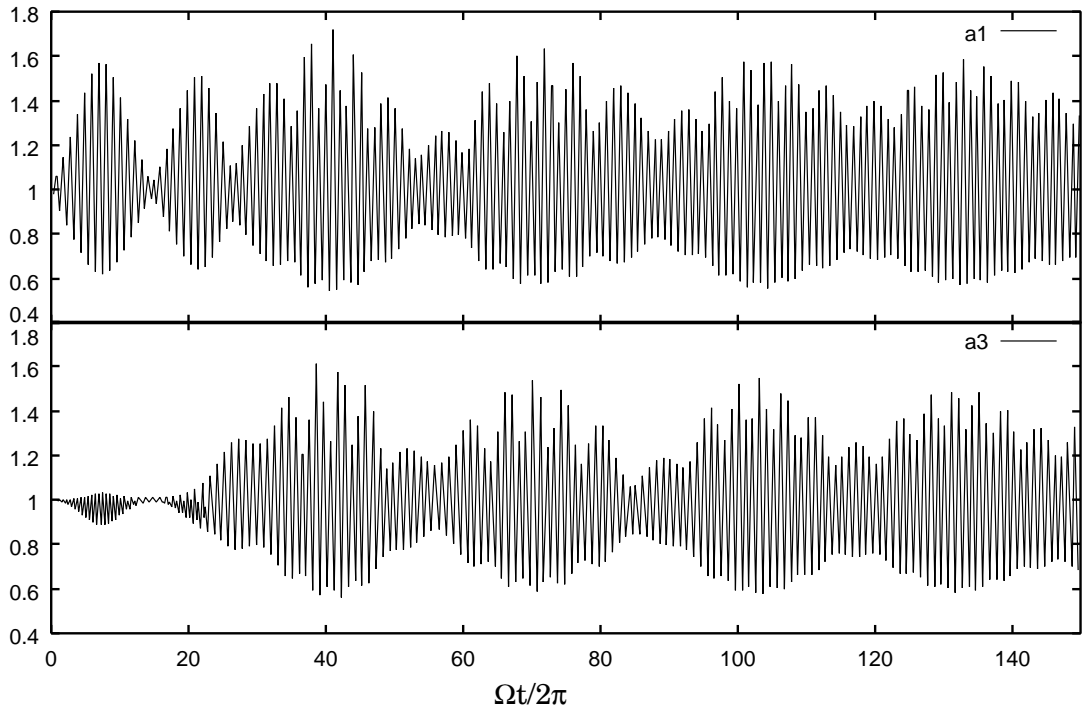}
\end{center}
\caption{
Oscillation amplitudes $\delta a_1$ and  $\delta a_3$
as a function of time $ \Omega t /(2\pi)$.
Parameters of the model are 
$ Q=100$, $ \omega =\Omega$ and  $h=5 \times 10^{-2}$ 
}  
\end{figure}

\section{ Summary and discussion }

  In this paper, we have examined resonant oscillations 
in ellipsoidal stellar model
with incoming gravitational wave.
Resonance occurs when a certain matching condition holds between
the stellar oscillation and the external periodic disturbance.
In the presence of radiation loss, 
the evolution significantly depends on 
a combination $ hQ = h  \nu  \tau $
consisted of the wave amplitude $h$, frequency $ \nu$ 
and the damping time $ \tau $.
The parametric resonance of the fundamental mode can grow up 
only if the quantity $ hQ  $ is larger
than a critical value $(\approx 7)$.
The amplitude of stellar oscillation increases, but
is saturated due to non-linear coupling,
by which energy is transfered to the additional oscillatory
motion. This point differs from linearized system, in which
the amplitude exponentially grows up.
If the quantity $ hQ  $ is smaller than the critical value,
the resonant oscillation is damped.
The resonant behavior at $  \omega =  \Omega $
is also determined by the parameter $ hQ  $.
If $ hQ  $ is larger than a critical value $(\approx 1)$,
then the system exhibits chaotic behavior.
We have also performed numerical calculations to explore
the possibility of parametric resonance in higher modes
$\omega/\Omega =n/2~(n=3,4,\cdots)$,
and found that the higher modes are
much more difficult to be excited.

In our idealized model, the external perturbation
is assumed to be constant.
This corresponds to the situation that 
duration $t_c$ of the driving source is much larger
than radiation damping time $\tau$, or
non-linear time scale $t_n$. 
If this condition does not hold, then the
environmental effect would appear first in the evolution. 
For example,  when the external perturbation stops at $t_c$,
the enhancement by the resonance also stops there,
and the amplitude is subsequently damped around $\tau (>t_c )$. 
  We now consider the astrophysical relevance of the resonant
oscillations.
The f-mode oscillation of a neutron star is
estimated as the frequency $ \nu =  $ a few kHz, 
the damping time $ \tau =$ a few $ \times 10^{-1}$ s, 
and therefore $ Q  \approx 10^2$.
In this case,  the exotic case needs large amplitude
of the gravitational wave 
$ h > Q^{-1} \approx 10 ^{-2}. $
The condition is realized only in hydrodynamical regime,
e.g, in core collapse of supernova and in the final phase of
binary coalescence.
The duration of periodic disturbance is also important factor.
Typically, the timescale is of order $10^{-3}$ s,
and may be too short for the mode to excited. 
Anyway, more realistic simulation is needed.

  Our result may be scaled to the excitation of the f-mode 
in white dwarfs. In this case, the frequency and damping time are 
given  by $ \nu \sim  0.1 $Hz, $ \tau  \sim 10^{11}$s, 
and therefore  $ Q \sim 10^{10}$.
In this case, the condition for the exotic case becomes
less severe $ h >  10 ^{-10}$.
The region available for this large amplitude 
may be estimated  by the total amount of
the radiated energy $\Delta M c^2 $.
The distance from the periodic gravitational wave source
is still limited to $ r \sim G \Delta M /(h c^2) \ll 1$pc.

\ack
This work was supported in part 
by the Grant-in-Aid for Scientific Research 
(No.14047215, No.16029207 and No.16540256) from 
the Japanese Ministry of Education, Culture, Sports,
Science and Technology.

\end{document}